# Coherent Population Trapping of an Electron Spin in a Single Negatively Charged Quantum Dot

Xiaodong Xu,[1] Bo Sun,[1] Paul R. Berman,[1] Duncan G. Steel,[1*] Allan S. Bracker,[2] Dan Gammon,[2] L. J. Sham[3]

[1] The H. M. Randall Laboratory of Physics,
The University of Michigan, Ann Arbor, Michigan 48109, USA
[2] The Naval Research Laboratory, Washington D.C. 20375, USA
[3] Department of Physics, The University of California-San Diego, La Jolla, California 92093, USA

[*] To whom correspondence should be addressed. E-mail: dst@umich.edu

**Coherent population trapping (CPT) refers to the steady-state trapping of population in a coherent superposition of two ground states which are coupled by coherent optical fields to an intermediate state in a three-level atomic system [1]. Recently, CPT has been observed in an ensemble of donor bound spins in GaAs [2] and in single nitrogen vacancy centers in diamond [3] by using a fluorescence technique. Here we report the demonstration of CPT of an electron spin in a single quantum dot (QD) charged with one electron. By applying a magnetic field in the Voigt geometry, we create a three-level lambda system, formed by two Zeeman sublevels of the electron spin and an intermediate trion state. As we tune the continuous wave driving and probe lasers to the two-photon Raman resonance, we observe a pronounced dip in the probe absorption spectrum that occurs because of the trapping of the electron spin population in the dark state. The observation demonstrates both the CPT of an electron spin and the successful generation of Raman coherence between the two spin ground states of the electron [4-6]. This technique can be used to initialize, at about GHz rate, an electron spin state in an arbitrary superposition by varying the ratio of the Rabi frequencies between the driving and probe fields. The results show the potential importance of charged quantum dots for a solid state approach to the implementation of**

**electromagnetically induced transparency (EIT) [7, 8], slow light [9], quantum information storage [10] and quantum repeaters [11, 12].**

When two radiation fields drive coupled transitions in a three-level lambda system, a steady-state coherent superposition of the ground states can be formed that is totally decoupled from the applied fields, a process that is referred to as CPT. A critical condition for realizing CPT is to have a pair of stable ground states with a relatively long coherence time compared to the excited state decay time. An electron spin trapped inside a single QD is a system that meets this requirement and constitutes an excellent opportunity for the realization of CPT. The demonstration of CPT shows the existence of the dark state which is important for various physical phenomena, e.g. CPT is the central physics of EIT [8], and the fast change of refractive index can lead to the effect of slow light [9], assuming an ensemble of identical charged quantum dots is available.

The electron spin inside a QD has been proposed as a qubit for quantum computing due to its long coherence time compared to fast optical operations [13]. An important step towards optically driven quantum computation in the QDs system is to generate electron spin coherence. The usual method of creating electron spin coherence in QDs is to use pulsed lasers [4-6]. Here, the demonstration of CPT by measurement of the absorption spectrum is evidence of the creation of electron spin coherence at a single QD level by continuous wave (CW) lasers.

Another critical element for quantum information science is the initial quantum state preparation [14]. Electron spin state initialization has recently been realized in a single QD by optical spin cooling techniques with a high fidelity [15, 16]. However, the limitation is that only two possible initial qubit states can be prepared, either spin up or spin down. CPT is a process that generates an arbitrary coherent superposition of electron spin ground states, whose probability amplitudes can be controlled by varying the ratio of Rabi frequencies between the driving and probe optical fields. Therefore, we can prepare an arbitrary initial qubit state by using the CPT technique. In this scheme, the initialization rate is limited by the excited state decay rate. In this system, the initialization rate is on the order of $10^9$ s$^{-1}$ [15].

The sample under study contains self-assembled InAs QDs embedded in a Schottky diode structure, which gives us the ability to control the charging state of the QD [15, 17]. In the experiment, we set the voltage such that only one electron is trapped inside the QD. At zero magnetic field, the energy level structure for the lowest lying states of the negatively charged QD can be modeled as shown in the inset on upper left of Fig. 1. The electron spin ground states are labeled as $\left|\pm\frac{1}{2}\right\rangle$, where $\pm\frac{1}{2}$ denotes the spin states along the growth axis. Since the trion states are formed by a spin singlet pair and one hole, the angular momentum of the trion state is determined by the hole spin and denoted as $\left|\pm\frac{3}{2}\right\rangle$. The only dipole allowed transition is from the spin ground state $\left|\frac{1}{2}\right\rangle$ ($\left|-\frac{1}{2}\right\rangle$) to the trion state $\left|\frac{3}{2}\right\rangle$ ($\left|-\frac{3}{2}\right\rangle$) with $\sigma^+$ ($\sigma^-$) polarized light excitation. Since the spin flip Raman transitions are dipole forbidden, the trion system at zero magnetic field can be considered as a double two-level structure, not possible for the realization of CPT.

In order to create a three-level lambda system, we apply a magnetic field in the Voigt geometry ($\vec{X}$ axis), i.e. perpendicular to the sample growth direction ($\vec{Z}$ axis). Since the electron and hole in-plane g factors are non-zero, the applied field mixes the spin ground states as well as the trion states. The energy level diagram and the associated selection rules of the trion system are shown in the inset at the upper right of Fig. 1. The new electron spin eigenstates $|X\pm\rangle$ can be excited to either trion states $|T\pm\rangle$ with linearly polarized light [15]. Hence, the forbidden Raman transitions at zero magnetic field are turned on when the magnetic field is applied along the $\vec{X}$ axis. As shown by the inset on the upper right of Fig. 1, we choose $|X\pm\rangle$ and $|T-\rangle$ to form a three-level lambda system. Since the hole g-factor is nonzero in InAs self-assembled QDs, the optical transitions with orthogonal polarizations from a trion state to the spin ground states are non-degenerate, thus suppressing the spontaneously generated coherence which was observed in GaAs interface fluctuation QDs [4].

We first characterized the QD with a single beam voltage modulation absorption experiment [15, 18]. We set the gate voltage at the edge of the trion charge plateau, where the optical pumping of the electron spin effect is suppressed [15, 16]. The main figure in Fig. 1 shows the quartet transition pattern of the trion state using $45°$ linearly polarized light to excite the transitions at a magnetic field of 1.32T. The observation of the four transition lines confirms that all four trion transitions are turned on and have similar transition strengths [15]. The four transitions are labeled as $V1$, $H1$, $H2$, and $V2$. The energy difference between $V2$ and $H1(H2)$ is the electron (hole) spin Zeeman splitting. We have studied the Zeeman splitting as a function of the magnetic field and obtained the electron and hole in plane g factors of 0.49 and 0.13, respectively, which are similar to the g factors reported in the literature [5, 15].

We then set the gate voltage to where the co-tunneling induced spin flip process is suppressed [19]. Figure 2(a) shows a single beam absorption spectrum by scanning the laser across transition H1 at a magnetic field of $2.64\ T$. We observed an almost flat line for the probe absorption spectrum reflecting the absence of the absorption due to optical pumping [15, 16]. The optical pumping induced saturation of the absorption shows that the spin relaxation rate is much slower than the trion relaxation rate. Hence, the spin ground states can be considered as meta-stable states compared to the short lived trion states.

To understand the experimental conditions for the measurements, we consider the interaction scheme shown in Fig. 3. A strong optical field (the driving field) is tuned on resonance with transition $V2$ and a weak optical field (the probe) is scanned across transition $H1$. When the probe laser is resonant with transition $H1$, the two-photon Raman resonance condition is reached. As seen in Fig. 2(b), a clear dip in the probe absorption spectrum is observed for $\Omega_d/2\pi = 0.56$ GHz. This observation demonstrates both the generation of the CPT of an electron spin and the Raman coherence between the spin ground states. For this particular set of data, the applied magnetic field is $2.64$ $T$, corresponding to an electron Zeeman splitting of $75.4$ $\mu eV$ ($18.2$ $GHz$).

The system is described by the optical Bloch equations for the three-level lambda system shown in Fig. 3, where $\Gamma_{ij}$ ($\gamma_{ij}$) is the trion population decay (dipole dephasing) rate, $\Gamma_s$ ($\gamma_s$) is the electron spin relaxation (decoherence) rate, $\Omega_i = \frac{\mu_i \times E_i}{\hbar}$ is the Rabi frequency, $\mu_i$ is the transition dipole moment, and $E_i$ is the optical field strength. For simplicity, we assume $\Gamma_s$, $\gamma_s \ll \Gamma_{ij}$, $\gamma_{ij}$, $\Omega_i$.

The analytical result of the probe absorption spectrum is generally complicated. However, a relatively simple form can be obtained when the driving and probe beams are both on resonance. After simplification, the absorption of the probe beam can be written as

$$\alpha = \alpha_o \frac{\left(\gamma_s - \Gamma_s + \lambda^2 \Gamma_s\right) \gamma_{T-X+}}{\left(1+\lambda^2\right)^2 \Omega_d^2}, \tag{1}$$

where $\lambda = \frac{\Omega_p}{\Omega_d}$ and $\alpha_o$ is a constant. In order to understand the physics of the dip, we can take $\lambda = 1$, the absorption expression is simplified to $\alpha = \frac{\alpha_o \gamma_{T-X+}}{4} \frac{\gamma_s}{\Omega_d^2}$. Therefore, the height of the dip is linearly proportional to the spin decoherence rate. If $\gamma_s = 0$ ($\gamma_s \ll \Omega_d$), the probe absorption vanishes (almost vanishes), *i.e.* the transition becomes transparent to the incident light due to the destructive interference of the coupled transitions driven by the coherent optical fields.

The observation of CPT can also be understood from the point of view of optical pumping. When the driving and probe lasers are on the two-photon Raman resonance, a coherent dark state is created that is decoupled from the applied optical fields and can be represented as $|D\rangle = \frac{\Omega_d |X+\rangle - \Omega_p |X-\rangle}{\sqrt{\Omega_d^2 + \Omega_p^2}}$. Part of the population is excited from the electron spin ground state to the trion state and relaxes spontaneously into the dark state. Since the dark state is not "seen" by the optical fields, the total population is eventually trapped there within a few radiative cycles of the trion state. In CPT, the coherence between the spin ground states is created by the coherent optical fields. Therefore, the whole process is an optical pumping process, whose rate is ultimately limited by the excited state decay rate, with the transfer of the mutual coherence between the optical fields to the electron spin coherence.

An arbitrary initial state for the quantum computation can be prepared by varying the ratio of the Rabi frequencies between the driving and probe fields. Ultimately, if we set $\Omega_d$ to zero, the initialized spin state will be $|X-\rangle$. This is the fast spin state preparation effect as discussed in Ref [15]. The difference is that when $\Omega_d$ is zero, there is no coherence involved in the state initialization, and the preparation efficiency is determined by the electron spin relaxation rate. In the initialization of the arbitrary coherent superposition state, we generate an electron spin coherence by the optical fields, and the state preparation efficiency is limited by the electron spin decoherence rate.

The linewidth of the dip in the probe absorption spectrum is ultimately limited by the electron spin decoherence rate. In the experiment, the smallest $\Omega_d$ we applied is $0.56$ $GHz$, which is about half of the trion transition linewidth, but still much larger than $\gamma_s$. Hence, the linewidth of the dip is broadened by the laser power. When $\Omega_d$ is strong, it will dress the spin ground state $|X-\rangle$ and the trion state $|T-\rangle$. In the case where $\Omega_d$ is larger than the trion transition linewidth, the absorption spectrum of the probe beam will split into two peaks when scanning across transition $H1$, which are known as Autler-Townes (AT) doublets [20], and has been demonstrated in a neutral QD [21, 22]. The spectral features of the probe absorption spectrum in our experiment is a combination of the AT splitting and the CPT quantum interference effect [7], where the spectral positions of the side bands can be determined by the AT splitting and the central feature in the absorption spectrum is due to the CPT effect, not a simple summation of the tails of the AT Lorentzian lineshapes.

The probe absorption spectra with various driving field and fixed probe Rabi frequencies are illustrated in Fig. 2(b-f). The energy separation of the two peaks is increased by increasing the driving field intensity. As $\Omega_d$ becomes larger than the trion transition linewidth, two AT peaks with Lorentzian lineshapes appear in the probe absorption spectrum, as shown by Fig. 2(e, f). Figure 4(a) displays the energy separation of the AT splitting peaks as a function of the driving field strength. A linear regression fits the data and extends to zero in the absence of the driving field, which indicates that the splitting is dominated by $\Omega_d$. The red solid lines on top of the data shown in Fig. 2(b-f) are the theoretical fits obtained by solving the optical bloch equations to all orders in the driving field and to first order in the probe. Assuming that $\gamma_s$ is a few orders of magnitude larger than $\Gamma_s$ (as we show below), we find $\gamma_{T-X+}/2\pi$, and $\gamma_s/2\pi$ equal to $(0.54\pm0.1)GHz$, and $(40\pm12)MHz$, respectively. The value of $40\ MHz$ corresponds to the electron spin decoherence time $T_2^*$ $(1/\gamma_s)$ of $4\ ns$. Although we measure an electron spin trapped inside a single QD, the electron spin $T_2^*$ extracted from the data is not the intrinsic electron spin decoherence time due to the hyperfine interaction between the electron spin and the neighboring nuclei ensemble [23-27]. The intrinsic $T_2$ can be measured by spin echo [28] or mode locking of spin coherence techniques [5].

The generation of the dark state is accompanied by the excitation of the electron spin coherence, which corresponds to the density matrix element $\rho_{x+x-}$. We inserted the parameters extracted from the fits into the optical Bloch equations and obtained values for the coherence between the spin ground states, which are represented by the red line in Fig. 4(b). The green line in Fig. 4(b) represents the theoretical values for the coherence in the absence of spin decoherence, given by $\frac{\Omega_d \Omega_p}{\Omega_d^2 + \Omega_p^2}$. The blue line represents the ratio of the experimentally generated coherence to the ideal case. The light blue dash vertical lines indicate the applied $\Omega_d$ in the experiment. At the maximally applied Rabi frequency 1.38 GHz, we infer that 94% of the optimal coherence is generated in our system.

Our results open the way to the demonstration of numerous quantum phenomena in spin based semiconductor QD systems. A direct step is to demonstrate electron spin ground state Rabi splitting by introducing a third CW laser, which is an analogy of electron spin Rabi oscillations in the time domain.

This work is supported by U.S. ARO, AFOSR, ONR, NSA/LPS, and FOCUS-NSF.

**Figure Captions**

**Fig. 1. The trion model and its characterization**. Inset: Trion energy level diagrams (upper left) without and (upper right) with magnetic field applied in the Voigt geometry. $V$ ($H$) means the transition is vertically (horizontally) polarized. At zero magnetic field, the spin flip Raman transitions are dipole forbidden. By applying a magnetic field in the Voigt geometry, the dark transitions become bright. A three-level lambda system is formed by these levels enclosed in the dashed line box. The main figure shows single beam absorption spectrum of the trion state at a magnetic field of 1.32 T with 45° linearly polarized light excitation. A quartet transition patterned is observed as the gate voltage is set in the non optical pumping region.

**Fig. 2. The experimental evidence of the CPT of an electron spin**. The gate voltage is set in the optical pumping region and the applied magnetic field is 2.64 T. (a) The probe absorption spectrum across transition H1 in the absence of the driving field. (b-f) Probe absorption spectra with various driving field Rabi frequencies. The driving field is set to be resonant with the transition from $|X-\rangle$ to $|T-\rangle$. The red solid lines are the theoretical fits by solving the optical Bloch equations. A pronounced dip is observed in the probe absorption due to generation of the dark state.

**Fig. 3**. **The interaction scheme of the generation of CPT.** A three level lambda system formed

by spin ground states $|X\pm\rangle$ of an electron and an intermediate trion state $|T-\rangle$.

**Fig. 4. The analysis of the CPT effect.** (a) The energy separation of the AT doublets as a function of driving field strength. (b) Theoretical curves of the creation of the electron spin coherence in a single charged quantum dot. Red line: experimentally generated electron spin Raman coherence ($\rho_{x+x-}$) inferred from the optical Bloch equations calculation by using the experimental parameters. The calculation is done under the experimental condition that the driving and probe fields are resonant with transition $V2$ and $H1$, respectively. Green line: the calculated maximum electron spin Raman coherence in the absence of the electron spin dephasing. Blue line: the ratio of the experimental generated coherence to the ideal case.

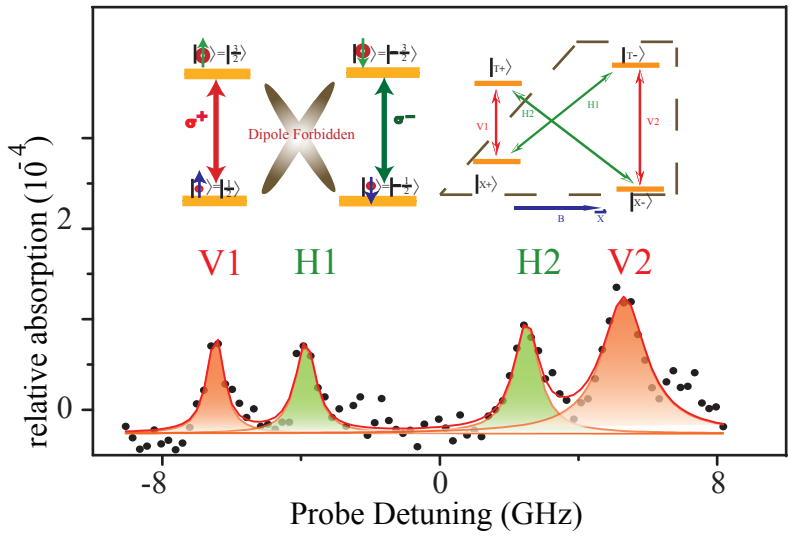

Figure 1

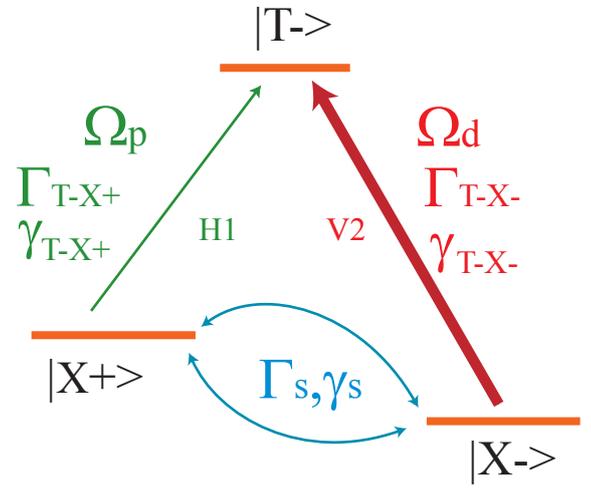

Figure 3

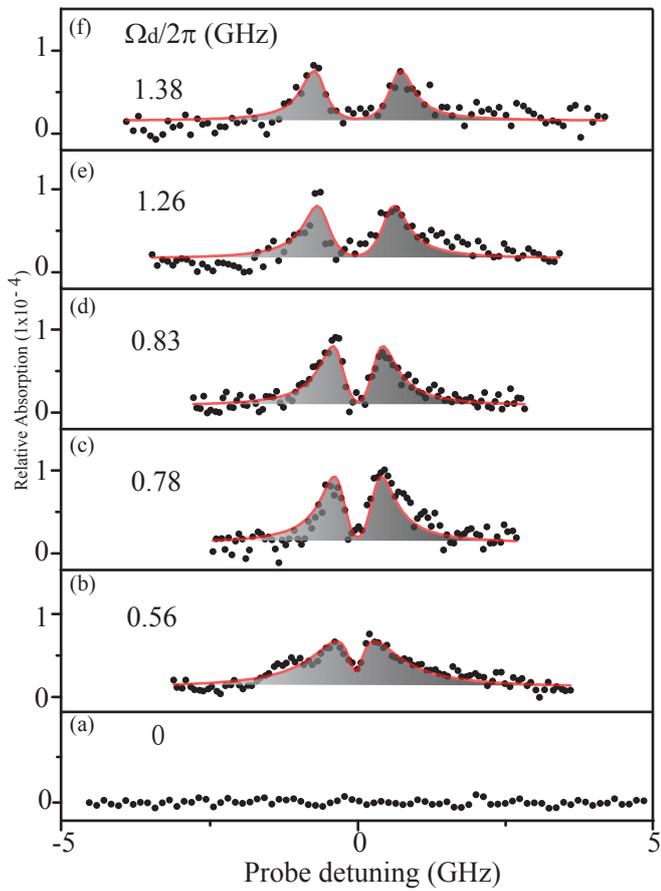

Figure 2

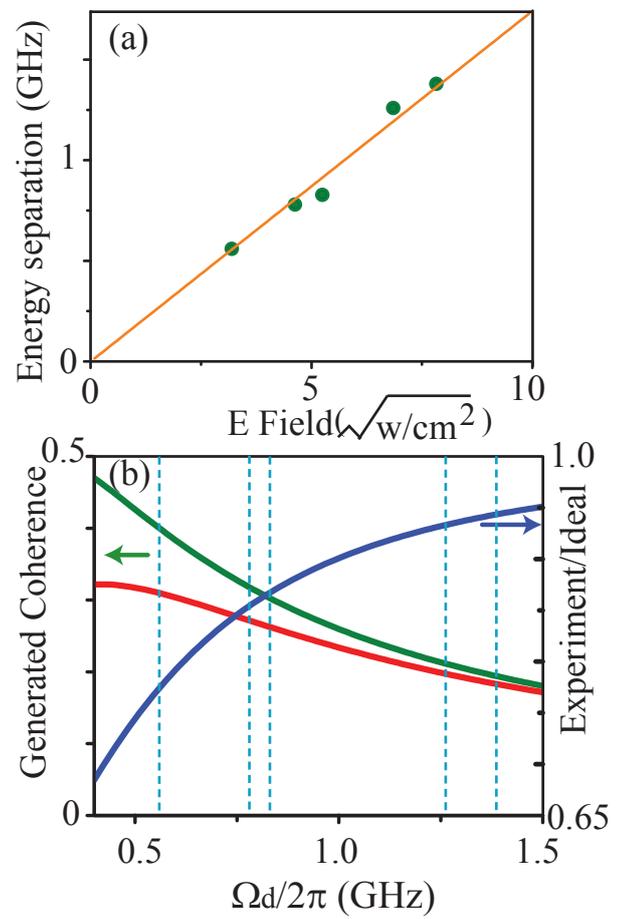

Figure 4